\documentclass[leqno]{article}

\usepackage{graphicx}  
\usepackage{verbatim}
\usepackage{subcaption}
\usepackage[letterpaper]{geometry}
\usepackage{algpseudocode}
\usepackage{bm}
\usepackage{amsmath}
\usepackage{todonotes}
\RequirePackage[OT1]{fontenc}
\RequirePackage[numbers]{natbib}

\newcommand{\ee}{{\Bbb E}}
\providecommand{\abs}[1]{\left\lvert#1\right\rvert}

\newcommand{\one}{{\hbox{1{\kern -0.35em}1}}}

\newcommand{\beq}[1]{\begin{eqnarray} \label{#1}}
\newcommand{\eeq}{\end{eqnarray}}
\newcommand{\bed}{\begin{displaymath}}
\newcommand{\eed}{\end{displaymath}}
\newcommand{\bea}{\bed\begin{array}{rl}}
\newcommand{\eea}{\end{array}\eed}
\newcommand{\disp}{\displaystyle}
\newcommand{\al}{\alpha}

\usepackage{ltexpprt}
\usepackage{hyperref}

\newtheorem{thm}{Theorem}[section]

\newtheorem{rem}[thm]{Remark}

\newtheorem{defn}[thm]{Definition}

\def\openbox{$\sqcup\llap{$\sqcap$}$}
\def\endproof{\unskip \enskip
    \null \nobreak \hfill \openbox \par}

\newcommand{\barray}{\begin{array}{ll}}
\newcommand{\earray}{\end{array}}

\usepackage{makeidx}  
\usepackage{amsfonts}
\makeindex            
\usepackage{times}  
\usepackage{graphicx} 

\begin{document}

\newcommand\relatedversion{}
\renewcommand\relatedversion{\thanks{The full python implementation  of the paper can be accessed at \protect\url{https://github.com/mpemy/optimaltradingml}}} 

\title{\Large Minimal Shortfall Strategies for Liquidation of a Basket of Stocks using Reinforcement Learning\relatedversion}
\author{Moustapha Pemy\thanks{Department of Mathematics, Towson
University, Towson, MD 21252-0001,
\ Email:
{\tt mpemy@towson.edu}}
\and Na Zhang\thanks{Department of Mathematics, Towson
University, Towson, MD 21252-0001,
\ Email:
{\tt nzhang@towson.edu}
}
}

\date{}

\maketitle

\begin{abstract} \small\baselineskip=9pt 
This paper studies the ubiquitous problem of liquidating large quantities of highly correlated stocks, a task frequently encountered by institutional investors and proprietary trading firms. Traditional methods in this setting suffer from the curse of dimensionality, making them impractical for high-dimensional problems. In this work, we propose a novel method based on stochastic optimal control to optimally tackle this complex multidimensional problem.
The proposed method minimizes the overall execution shortfall of highly correlated stocks using a reinforcement learning approach. We rigorously establish the convergence of our optimal trading strategy and present an implementation of our algorithm using intra-day market data.

\end{abstract}


\section{Introduction}
Algorithmic trading has been the primary trading methodology in financial markets over the past three decades. Investors of all sizes in the exchange market rely on various types or methods of algorithmic trading to optimize the efficiency and outcomes of their executions. This is especially crucial for investors engaged in high-frequency, large-volume trading of multiple highly correlated assets on a daily basis. For these participants, maximizing efficiency is essential as inefficiency can result in substantial losses. This question has been extensively investigated in the literature.  However, the overwhelming approaches employed fail to address this problem efficiently in high-dimensional settings because of the curse of dimensionality. Thus, it is crucial to devise an efficient liquidation strategy for a basket of highly correlated stocks and ensure it overcomes the curse of dimensionality.

The execution problem in a multidimensional setting has been explored in the literature. Here, we would like to highlight some recent works. Angoshtari and Leung (2020) proposed using a multi-dimensional Brownian bridge to model the joint dynamics of futures prices and solve stochastic optimal trading problems, addressing both scenarios where the underlying assets are traded and where they are not.   Yang et al. (2020) apply reinforcement learning to optimize stock trading strategies, proposing an ensemble stock trading strategy that combines Proximal Policy Optimization, Advantage Actor-Critic, and Deep Deterministic Policy Gradient to maximize investment returns.  Recent 
contributions by Sun et al. (2023), Pemy and Zhang (2023), and Pemy and Zhang
(2024)  extend these ideas further into machine learning applications in finance.

In this paper, unlike the problems studied in the aforementioned references, which primarily focus on trading strategy optimization or stochastic modeling of price dynamics, we specifically aim to minimize the overall execution shortfall for a basket of stocks. Our focus is on addressing the critical issue of execution efficiency in multi-stock trading, ensuring minimal deviation between theoretical and actual trade outcomes.
The main issue when trading a basket of stocks is fully integrating the advantages and disadvantages of correlation. Any rigorous trading strategy should optimally account for all aspects of correlation and its influence on asset volatility. Whenever classical methodologies have attempted to meaningfully integrate correlation, this has led to an intractable problem, resulting in the well-known curse of dimensionality. Linear quadratic regulator methods have failed miserably when tackling multidimensional highly correlated complex problems. To address this challenge, we propose a novel algorithm based on reinforcement learning which will optimally tackle the multidimensional cases. Our method is inherently based on a multidimensional dataset. As such, high dimensionality is not a hindrance but a great advantage of our methodology. We train a set of neural networks to learn the intricacies and interdependencies among multiple stocks. As a matter of fact,  we assume without loss of generality that those stocks are correlated and we will use their correlation to our advantage. We study this problem as a discrete-time optimal control problem. More precisely, we use the reinforcement learning methodology to tackle this problem. We propose a minimal executive shortfall algorithm in which we train two neural networks  simultaneously to understand the nuances of the stocks involved so as to miminize their overall implementation shorfall. The environment in which we train our neural networks is a discrete-time controlled dynamical system where stock prices follow a discrete version of the geometric Brownian motion, and our trading actions constitute the control variables.
 
The remainder of the paper is organized as follows. 
After presenting the problem at hand in Section 2, we review relevant definitions. In Section 3, we propose a minimal shortfall algorithm and prove its convergence. In Section 4, we present the implementation of the algorithm and in Section 5, we outline an application of our method using real market data.

\bigskip

\section{Trading Model} 
Consider an institutional investor who wants to re-balance its portfolio thus needs to liquidate large blocks of shares of $n$ stocks  $S_1, S_2, \dots, S_{n}$.  Moreover, we assume that these stocks $S_1, S_2, \dots, S_n$, are all highly correlated.  The total number of shares of the $i$-th stock we have to liquidate is denoted by $N_i<\infty$.   In addition, the entire execution should be completed within a short time horizon, say $0<T<\infty$. Our main goal is to tackle the optimal liquidation problem of highly correlated stocks using a reinforcement learning approach. 
Let $S_{i,k}$ be the share price of the $i$-th stock at the $k$-th trading period respectively for $i=1, \dots, n$ and $k\geq 0$.  Since our main goal is to sell those stocks as efficiently as possible, we denote by  $a_{i,k}$  the number of shares sold or bought of the $i$-th stock at the $k$-th trading period respectively for $i=1, \dots, n$ and $k\geq 0$. Note that the quantities $a_{i,k}$ represent the actions of the trader, in other terms, the control variables. 
Let the functions $f_i: {\Bbb R} \to {\Bbb R}$, and the constants  $\mu_i$, and $\sigma_{ij}$ represent respectively the relative impacts,  rates of return and covariances of these stocks for $i,j = 1, \dots, n$.  We assume that the stock prices dynamics are captured by an operator $H$; for vectors $S_{k} = (S_{1,k}, \ldots, S_{n,k})$, $\xi_k=(\xi_{1,k}, \ldots, \xi_{n,k})$ and $a_k = (a_{1,k}, \dots, a_{n,k})$, we have $ S_{k+1} = H(S_k, a_k,\xi_k)$ such that
\beq{dynamics} \label{eq: model}
\qquad \left\{\begin{array}{ll}
\disp S_{i,k+1} &=  S_{i,k} + \tau \big(f_i( a_{i,k})+ \mu_i \big)S_{i, k} \\
&\disp +  \sqrt{\tau} \sum_{i,j=1}^n \sigma_{ij} \xi_{j,k}S_{i,k},\\
S_{i,0} &= s_i >0, \quad i= 1,..., n; \quad  k\geq 0
 \end{array}\right.
\eeq
where $(\xi_{i,k})_k$, $i=1,..., n$, are sequences of independent and identically distributed random variables, the quantity $\tau$ is the time unit, it can roughly be seen as the smallest time elapsed between two consecutive trades.  The random variables $\xi_{i,k}$, $k=1,2,..., i=1, ...,n$ are independent and normally distributed random variables with mean 0 and variance 1. We assume that the processes $(\alpha_k)_k$ and $(\xi_k)_k$ are independent.  All processes and random variables are defined on a given probability space $(\Omega, {\cal F}, {\Bbb P})$.\\
 The functions $f_i$  represent the relative impact on the share price of trading a certain number of shares. In fact, if we frequently put large blocks of shares in the market this will definitely bring the price down. In order to model that effect we have taken a linear function. Thus, $f_i(x)=\lambda_i x$, $\lambda_i\in(-1,1)$, with $\lambda_i <0$ when selling and  $\lambda_i >0 $ when buying.
\begin{defn}\label{policy}
\begin{itemize}
    \item[1.] A trading strategy or policy is a finite sequence of vectors $\pi =(a_k)_k= ( a_{i,k})_{1\leq i\leq n,k \geq 0}$, such that $\disp \sum_{k=0}^\infty a_{i,k}=N_i$. Note that the $a_{i,k}$  are zeros except for a few of them.
    \item[2.]  Let $V_{i,k}$ be the volume of the  $i$-th stock traded at the $k$-th trading period. The market volume-weighted price (VWAP) for the $i$-th stock after $k$ trades is defined as follows
\beq{marketVwap}
M_{i,k} \disp =  \frac{\sum_{j=1}^k V_{i,j} S_{i,j}}{\sum_{j=1}^kV_{i,j}},
\eeq
and the trader volume-weighted price (VWAP) of the $i$-th stock after $k$ trades is defined as
\beq{traderVwap}
T_{i,k} \disp =  \frac{\sum_{j=1}^k a_{i,j} S_{i,j}}{\sum_{j=1}^ka_{i,j}}.
\eeq
\end{itemize}
 
\end{defn}

\section{The Minimal Shortfall Execution}
Minimum shortfall methods are usually applied to liquidate orders. In this section, we propose a liquidation strategy that addresses the issue for a basket of stocks.
The main goal of liquidating orders is to minimize the shortfall.

 Consider a basket of stocks $(S_{1,k})_k, (S_{2,k})_k, ..., (S_{n,k})_k$  defined in \eqref{dynamics}. For each stock $(S_{i,k})_k, i=1,..., n$, let $\bar{s}_i$ represents the arrival price for the $i$-th stock, i.e., the stock price value when the execution order is given. The  expected shortfall of the $i$-th stock at till the $k$-th trading time is defined as follows
 \beq{shortfalli}
F_{i,k} = \sum_{j=1}^k  \mathbb{E}\bigg[a_{i,j}(S_{i,j}  -  \bar{s}_i)\bigg] 
\eeq
where $\mathbb{E}$ represents the probability expectation,
and the overall expected shortfall for the basket of stocks after the $k$-th trading time  is 
\beq{shortfall}
F_k = \sum_{i=1}^n F_{i,k}.
\eeq
Our goal is to devise a trading algorithm that will over a finite horizon minimize the process $(F_k)_k$. We will search for a strategy such that the stock $S_i$ is sold as close as possible to the arrival price $\bar{s}_{i}$, $i=1, \ldots, n$. Let $x_{i,k}:= S_{i,k}-\bar{s}_i$ be the shortfall tracking error for the $i$-th stock at the $k$-th trading time.  In other words, our goal is to drive the error $x_{i,k}$ as close as possible to zero as we are selling for each $i=1, \ldots, n$.
Since we assume the stock prices dynamics are captured by the operator $H$, given by\eqref{eq: model}, denote the shortfall error vector by $x_k=(x_{1,k}, \ldots, x_{n,k})$ and the arrival price vector by $\bar{s}=(\bar{s}_1, \ldots, \bar{s}_n)$, it is easy to see that $x_{k+1}=H(x_k,a_k, \xi_k)$ satisfying

\beq{error dynamics} \label{eq: error model}
\qquad \left\{\begin{array}{ll}
\disp x_{i,k+1}&=x_{i,k} + \tau \big(f_i( a_{i,k})+ \mu_i \big)x_{i, k} +  \sqrt{\tau} \sum_{i,j=1}^n \sigma_{ij} \xi_{j,k}x_{i,k}\\
&+\bar{s}_i\big(f_i( a_{i,k})+ \mu_i \big)+\bar{s}_i\sqrt{\tau} \sum_{i,j=1}^n \sigma_{ij} \xi_{j,k},\\
x_{i,0} &= s_i-\bar{s}_i, \quad i= 1,..., n; \quad  k\geq 0.
 \end{array}\right.
\eeq

Then the expected shortfall can be written as a function of the tracking error:

 \beq{shortfalli_trackingerror}
F_{i,k} = \sum_{j=1}^k  \mathbb{E}\bigg[a_{i,j}x_{i,k} \bigg], \text{ and } \ F_k = \sum_{i=1}^n F_{i,k}. 
\eeq

To achieve this goal, we propose a new computational method, i.e.,  a reinforcement learning-based linear quadratic regulator approach. We will first define the $Q$-function for the minimal shortfall execution. 

\begin{defn}\label{q-function1}
 Given a policy $\pi$ with initial action $a_0=a$ and given an initial tracking error $x_0 =x$ with  $x= s-\bar{s}$ 
 we define the $Q$-functions of the policy $\pi$ as follows
 \beq{value_function_iterative}
 Q^\pi_p(x, a)  \disp = \ee\bigg[ \sum_{k=1}^p \sum_{i=1}^n  \gamma^{k-1}F_{i,k}
  \bigg{|} e_0=x, a_0=a; \pi\bigg],
 \eeq
 \beq{value-function}
 Q^\pi(x, a)\disp = \ee\bigg[ \sum_{k=1}^\infty \sum_{i=1}^n  \gamma^{k-1}F_{i,k} 
   \bigg{|} e_0=x, a_0=a; \pi\bigg].
 \eeq 
 One can obviously see that $\disp \lim_{p\to \infty } Q^\pi_p(x, a)  = Q^\pi(x, a) $.\\
The state-action value functions of our control problem are defined as follows
 \beq{valuefunctioniterative1}
Q_p(x,a) \disp= \min_{\pi}Q^\pi_p(x,a)
\eeq
 and
\beq{valuefunction1}
Q(x,a) \disp= \min_{\pi}Q^\pi(x,a).
\eeq
We have $\disp \lim_{p\to \infty } Q_p(x, a)  = Q(x,  a) $.
\end{defn}

Denote the state space of $(x_k)_k$ by $\mathcal{S}$. Our main goal is, for each $x\in\cal S$,   to find the optimal policy, say $\pi^*$, such that 
\beq{prob1}
Q(x, a) \disp= \min_{\pi}Q^\pi(x, a) = Q^{\pi^*}(x, a).
\eeq

\subsection{A Reinforcement Learning Algorithm}

It is easy to see that in order to solve \eqref{prob1}, 
 we need to use the famous Bellman Dynamical Programming Principle equation
 
 \beq{DDPshortfall}
Q_{p+1}(x, a) = \ee\bigg[F_{p}  + \gamma \min_{\pi} Q_{p}(x', a')
 \bigg{|} x'=x, a'=a \bigg], \nonumber \\
  p=0, 1, \cdots.
\eeq
Theoretically, the iterative value function $Q_{p+1}(x, a) $ converges to the value function $Q(x, a).$ But this approach usually leads to the curse of dimensionality and many other stability issues. In order to address these issues, 
we will use parametric function approximations. 
In this methodology we will simultaneously approximate the optimal policy and the value function.  Each parametric function approximator is a neural network. Therefore, we will have two neural networks. The weights of our neural network will be the parameters of our function approximators. 

Denoted by $\cal{A}$ the state space of actions $(a_k)_k$, we call it the action space. It is clear that the triplet$(x_k, a_k, R_k)_k$ constitutes a Markov Decision process with transition probabilities $P(x, a,  x'):={\Bbb P}(x_{k+1}=x', |x_k=x, a_k=a)$, for any $x,x'$  in state space ${\cal S}$ and  $a$ in action space ${\cal A}$. 
We define our policy function as follows $\pi(x, a, \theta):= {\Bbb P}(a_k = a|x_k = x,\theta)$, $\forall x\in\cal S$, $a\in\cal A$, where $\theta$ is the vector of parameters of the neural network that will approximate the policy. For simplicity, we write just $\pi(x, a )$ for $\pi(x, a,  \theta)$.
To solve this problem we make the following assumptions:
\begin{itemize}
\item {\bf (A1)}: We assume that under any policy $\pi=(a_k)_k$ our Markov Decision Process is irreducible and aperiodic.

\item {\bf (A2)}: We assume that our policy function $\pi(x, a, \theta)$ is differentiable with respect to $\theta$.
\end{itemize}

We define the average reward as 
\beq{averageReward-shortfall}
\rho(\pi) = \ee\bigg[\sum_{k=1}^\infty\gamma^{k-1}F_k \bigg{|} x_0,  \pi \bigg]
\eeq
and  we have
\beq{QfunctionII-shortfall}
Q^{\pi}(x, a) = \ee\bigg[\sum_{k=1}^\infty\gamma^{k-1}F_k 
\bigg{|} x_0=x,a_0 = a, \pi \bigg].
\eeq

Moreover we define the discounted states visited starting in state $s_0$ and following the policy $\pi$ as following
\begin{equation}\label{strating}
d^{\pi}(x)= \sum_{k=0}^\infty \gamma^k {\Bbb P}(x_k = x| x_0,  \pi).
\end{equation}  

It  is well known from the literature Sutton et al. \cite{Sutton2000}, and Marbach and Tsitsiklis \cite{Marbach2001} that,  we have 
\beq{Policy-shortfall}
\qquad\,\,\,\,\nabla_{\theta} \rho(\pi)
= \sum_{x\in\cal S}d^\pi(x) \sum_a \nabla_{\theta} \pi(x, a, \theta)Q^\pi(x, a).
\eeq

It is clear that anytime the policy $\pi$ is updated this also affects the state-action value function $Q^\pi(x, a)$. Thus given that we will approximate the policy with a parametric function (neural network), we will also approximate the state-action value function with another parametric function approximator denoted $\hat{Q}_\omega^\pi(x, a)$ where $\omega$ represents the vector of weights of the second neural network. One of the necessary conditions for the convergence of the approximation of $Q^\pi$ is that its approximator minimizes the mean-squared error:
\beq{error}
\qquad e^\pi(\omega)&=& \sum_{x \in {\cal S}}d^\pi(x)\sum_{a {\cal A}}\pi(x,a)\bigg( Q^\pi(x,a)  \\
&&\hspace{0.5 in} - \hat{Q}^\pi_\omega(x,a)\bigg)^2, \nonumber
\eeq
which means that we should at least have
\beq{condition}
\nabla_\omega e^\pi(\omega) &=&  2 \sum_{x \in {\cal S}}d^\pi(x)\sum_{a {\cal A}}\pi(x,a)\bigg( Q^\pi(x,a)  \nonumber\\
&&\hspace{0.5 in} - \hat{Q}^\pi_\omega(x,a)\bigg)\nabla_\omega \hat{Q}^\pi_\omega(x,a) \\
&=&0. \nonumber
\eeq
Sutton et al. \cite{Sutton2000} showed any if an approximator $\hat{Q}^\pi_\omega(x,a)$ satisfies \ref{condition} and is such that
\beq{4Eq} 
 \disp \nabla_\omega \hat{Q}^\pi_\omega(x,a)= \frac{\nabla_\theta\pi(x,a)}{\pi(x,a)}
\eeq then
\beq{functionGradient}
\qquad \qquad \nabla_\theta \rho^\pi(\theta) =\sum_{x \in {\cal S}}d^\pi(x)\sum_{a\in {\cal A}}\nabla_\theta\pi(x,a)\hat{Q}^\pi_\omega(x,a).
\eeq
Moreover, we also assume that our policy function $\pi(x,a,\theta)$ is in  the form of a Gibbs (Boltzmann) distribution
\beq{policyform}
\quad \quad \quad\disp \pi(x,a, \theta) = \frac{e^{\theta^T\phi_{xa}}}{\sum_{a'}e^{\theta^T\phi_{xa'}}}, \quad \forall x \in {\cal S},\,\,\, \forall a \in {\cal A},
\eeq
where $\phi_{xa}$ is a feature vector characterizing the state-action pair for each $x$ and $a$. So using \ref{4Eq} and \ref{policyform} we have
\beq{NewEq}
\nabla_\omega\hat{Q}^\pi_\omega(x,a)= \phi_{xa}-\sum_{b\in{\cal A}} \pi(x,b)\phi_{xb}.
\eeq
 Since it is preferable for  $\hat{Q}^\pi_\omega(x, a)$ to be a linear function of $\omega$ in order to satisfy all convergence conditions, thus this implies that 
 \beq{MousEq}
\qquad \qquad \hat{Q}^\pi_\omega(x,a) = \omega^T\bigg(  \phi_{xa}-\sum_{b\in{\cal A}} \pi(x,b)\phi_{xb} \bigg).
 \eeq 

\begin{theorem}
Given any initial parameter vector $\theta_0$,  the sequence of average reward $(\rho(\pi_k))_{k \in \mathbb{N}}$, with $\pi_k=\pi(\cdot, \cdot, \theta_k)$ and 
\beq{wsed}
\omega_k \disp= \omega \quad \hbox{ such that }  \quad \nabla_\omega e^{\pi_k}(\omega) = 0
\eeq
and
\bea
 \theta_{k+1}&\disp  = \theta_k + \frac{1}{k}\sum_{x\in {\cal S}}d^{\pi_k }(x)\sum_{a\in {\cal A}}\nabla_\theta\pi_k(x, a) \hat{Q}^{\pi_k}(x, a)
\eea is such that 
\beq{lim}
\lim_{k\to \infty} \nabla_{\theta}\rho(\pi_k) = 0.
\eeq   
\end{theorem}

\paragraph{Proof.}
Given the total number of shares that we want to liquidate is finite $\sum_{i=1}^n N_i<\infty$, then the states that our MDP will visit and the actions it will take until we complete our liquidation are both finite. So the sets ${\cal E}$ and ${\cal A}$ are finite sets. Using \eqref{policyform} it is easy to see that when $a$ and $x$ are fixed, the second partial derivative of the policy function $\pi(x, a,  \theta)$ with respect to theta is bounded. In fact, we have 
\beq{derivatives}
\abs{\frac{\partial^2 \pi(x, a,  \theta)}{\partial \theta_i\partial \theta_j}}< B_{x, a} \qquad \forall i, \, j.
\eeq
Moreover, given that the state spaces ${\cal S}$, and the action space ${\cal A}$  are finite sets then there exists $B<\infty$  such that
\beq{maxDeriv}
\max_{\theta, x, 
a, i, j}\abs{\frac{\partial^2 \pi(x,a, \theta)}{\partial \theta_i\partial \theta_j}}< B<\infty.
\eeq
Moreover our shortfall $ F=(F_k)$ is uniformly bounded because after a while the trader will have finished the liquidation and exited the market, so there exits $K>0$ such that for any $n >K$, $F_n=0$. It is clear that the sampling step-size sequence $\beta_k = \frac{1}{k}$ satisfies the condition
\bea
&\disp \sum_{k}\alpha_k = \sum_{k}\frac{1}{k} = \infty \\
 &$ and $ \lim_{k\to \infty} \beta_k = \lim_{k \to \infty} \frac{1}{k}=0.
\eea
Consequently using Theorem 3 from  Sutton et al. \cite{Sutton2000} we have have the convergence of the sequence of average reward in the sense that 
\bea
\disp \lim_{k \to \infty} \nabla_\theta \rho(\pi_k) = 0.
\eea
\endproof.
\section{Implementation}

The training environment of our reinforcement learning schema is our model \ref{dynamics}. In this model we assume the stocks we want to liquidate are highly correlated and that our trading actions will incur a linear market impact.  We store experimental data $\epsilon_k = (x_k, a_k, F_k, x_{k+1})$ at the $k$-th step in a   data set $ D_k = \{\epsilon_1, \epsilon_2,\ldots, \epsilon_k\}$. During learning, we estimate Q-function by drawing uniformly at random from the pool of stored data. We use the following algorithm.
\begin{algorithm}
\begin{algorithmic}\\
\State Initialize replay memory $D$ to capacity $N_D$.
\State Initialize the action-value function estimate $\hat{Q}_\omega$  with random weights $\omega$.  
\State Initialize the policy function $\pi(\cdot, \cdot, \theta)$  with random weights $\theta$
\State Initialize the target network  $\hat{Q}'_{\omega'}$ and $\pi'(\cdot, \cdot, \theta')$ with weights $\omega' \gets \omega$ and $\theta' \gets \theta$.
\For{ episode=1 to $M$ }\\
    Received the first state $e_1$
     \For{k=1 to K}
    \State  With probability $\varepsilon$ select a random action $a_k$ otherwise
      set  $a_k = \arg\max_{a \in {\cal A}}\hat{Q}_\omega(x_k, a, \omega)$
     \State  Execute action $a_k$ and observed reward $R_k$, then compute the next state $x_{k+1} = H(x_k, a_k,\xi_k)$.
	\State Store transition $\epsilon_k=(x_k, a_k, F_k, x_{k+1})$ in $D$.
	\State Sample a random mini-batch of $N_D$ $(x_i,a_i, F_i, x_{i+1})$  transitions from $D$.
	\State Set $y_i = F_i +\gamma \hat{Q}_{\omega'}'(x_{i+1}, a_{i+1})$.
	\State Update  $\omega_{k+1}$ so as to minimize the estimated mean-squared error $ L= \frac{1}{N_D}\sum_i(y_i-\hat{Q}_\omega(x_i,a_i))^2$ using gradient descent.
  \State Update policy gradient using the sample gradient $\nabla_\theta \rho \approx \frac{1}{N_D}\sum_{i} \nabla_\theta \pi(x_i,a_i) \hat{Q}_\omega(x_i,a_i) $
\State $\theta_{k+1}  =  \theta_{k} + \frac{1}{k}\nabla_\theta \rho$
    \State Every C steps, reset $\hat{Q}' \gets \hat{Q}$,  $\pi' \gets \pi$, $ \omega'\gets \omega$ and $\theta' \gets \theta$. 
      \EndFor
\EndFor

\end{algorithmic}
\end{algorithm}

\section{Application}

In this section,  we test our model with a set of intraday trading data from six highly traded and correlated stocks: Apple (AAPL), Google (GOOG), IBM (IBM), ATT (T), Verizon (VZ),  and Exxon Mobil (XOM) during the week of September 5 to 8, 2017.  Using the standard statistical estimation methodology we calibrate the model \eqref{dynamics}, the expected mean returns of these stocks in September are
\bea
\mu_1 = -3.53\times 10^{-6} ,\quad \mu_2 = -2.16\times 10^{-7}\\
\mu_3 = -1.49\times 10^{-6} ,\quad \mu_4 = -1.06\times 10^{-5}\\
\mu_5 = -7.66\times 10^{-6} ,\quad \mu_6 = 5.94\times 10^{-6},
\eea
where $i=1, 2, ...,6$ represent AAPL, GOOG, IBM, T, VZ, and XOM respectively. Their covariance matrix $(\sigma_{ij})_{1\leq i,j\leq 6}$ for September   is
\bea
&\left[ \begin{array}{llllll}
6.5 & 0.17 & 0.03 & -0.015 & 0.26 & 0.15\\ 
0.17 & 6.9 & -0.088 & 0.15 & -0.012 & 0.045\\
0.03 & -0.088 & 4.46 & 0.007 & 0.065 & 0.37\\
-0.015& 0.15 &0.007 &6.98& 0.011& 0.019\\
0.26&-0.12&0.065&0.011&6.46&0.13\\
0.15& 0.045& 0.37&0.019& 0.13 &5.7
\end{array}
\right]\\
&\quad\times 10^{-8}.
\eea 

 For each of these stock we have a threshold $M_i$ that we can traded at once without been detected. In fact, $M_i$ represents the maximum number of shares we can sell or buy at the given traded for the $i$-th stock. Consequently the action $a_{i,k} \in [-M_i, M_i]$.  In this particular case we have
\bea
M_1 =200, \quad M_2 =50\\
M_3 = 200, \quad M_4 = 500\\
M_5 = 500, \quad M_6 = 200.
\eea
Moreover, the market impact functions are $f_i(x) = \lambda_i x $ with $\lambda_i = -1\times 10^{-11}$, $i=1,...,6$.  
For the week of September 5 to 8, 2017, the trading data of the first three days are used as training data and the trading data of the last day of the week are the testing data. Figures 1 and 2 show the expected tracking error and the expected shortfall for each of these stocks during the last day of each of these periods. Figure 3 illustrates the expected tracking error for each stock based on specific actions taken. We can see that our methodology 
clearly achieves its intended objectives and can be effectively implemented in a software package.


\begin{figure*}
        \caption[  The Tracking Errors for Apple (AAPL), Google (GOOG),  IBM (IBM), ATT (T), Verizon (VZ), and Exxon Mobil (XOM) on Sept. 8, 2017]
        {\small The Tracking Errors for Apple, Google,  IBM, ATT, Verizon, and Exxon Mobil on September 8, 2017} 
        \label{fig: mean and std of nets}
        \centering
        \begin{subfigure}[b]{0.45\textwidth}
            \centering
            \includegraphics[width=\textwidth]{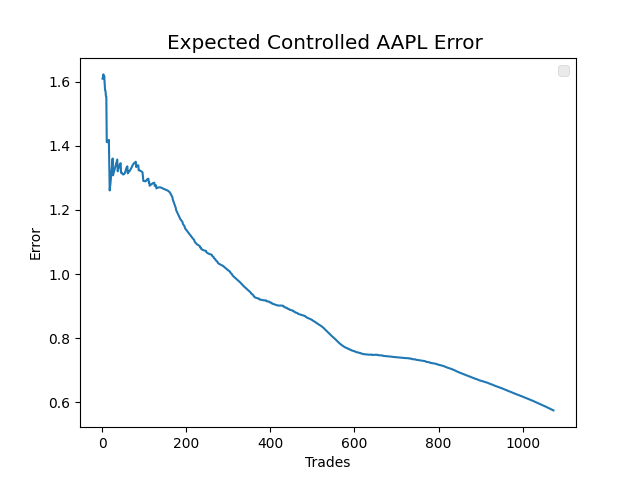}
            \caption[Network2]%
            {{\small Apple Controlled Error on September 8, 2017 }}    
            \label{fig:mean and std of net14}
        \end{subfigure}
        \hfill
        \begin{subfigure}[b]{0.45\textwidth}  
            \centering 
            \includegraphics[width=\textwidth]{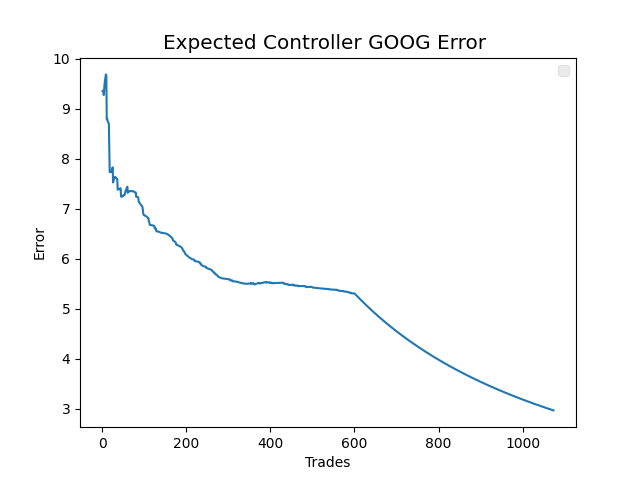}
            \caption[]%
            {{\small Google Controlled Error on September 8, 2017}}    
            \label{fig:mean and std of net24}
        \end{subfigure}
        \vskip\baselineskip
        \begin{subfigure}[b]{0.45\textwidth}
            \centering
            \includegraphics[width=\textwidth]{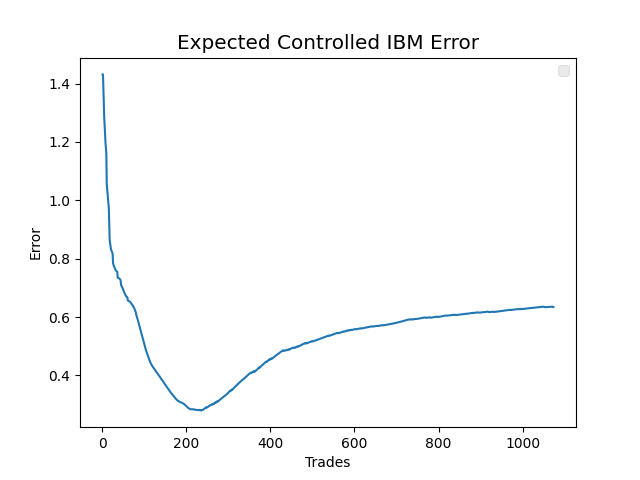}
            \caption[Network2]%
            {{\small  IBM Controlled Error on September 8, 2017}}    
            \label{fig:mean and std of net34}
        \end{subfigure}
        \hfill
        \begin{subfigure}[b]{0.45\textwidth}  
            \centering 
            \includegraphics[width=\textwidth]{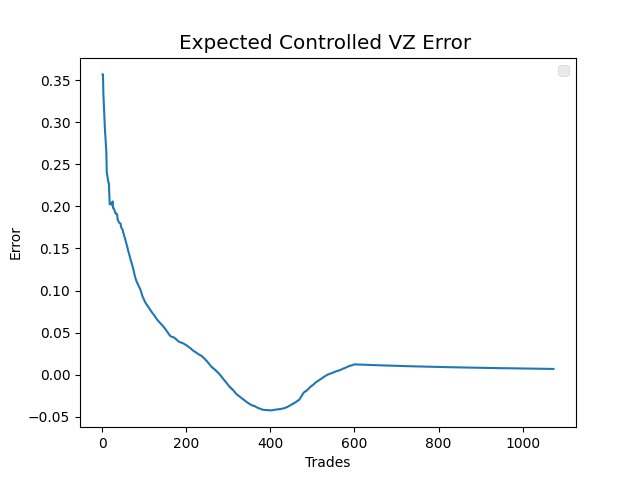}
            \caption[]%
            {{\small  VZ Controlled Error on September 8, 2017}}    
            \label{fig:mean and std of net44}
        \end{subfigure}
        \vskip\baselineskip
        \begin{subfigure}[b]{0.45\textwidth}   
            \centering 
            \includegraphics[width=\textwidth]{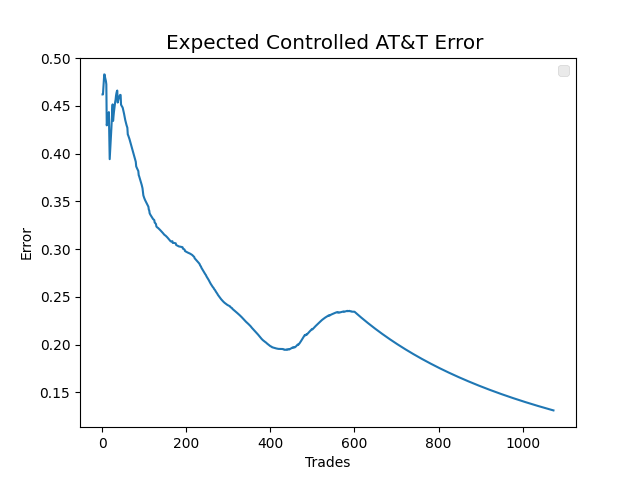}
            \caption[]%
            {{\small ATT Controlled Error on September 8, 2017}}    
            \label{fig:mean and std of net54}
        \end{subfigure}
        \hfill
        \begin{subfigure}[b]{0.45\textwidth}   
            \centering 
            \includegraphics[width=\textwidth]{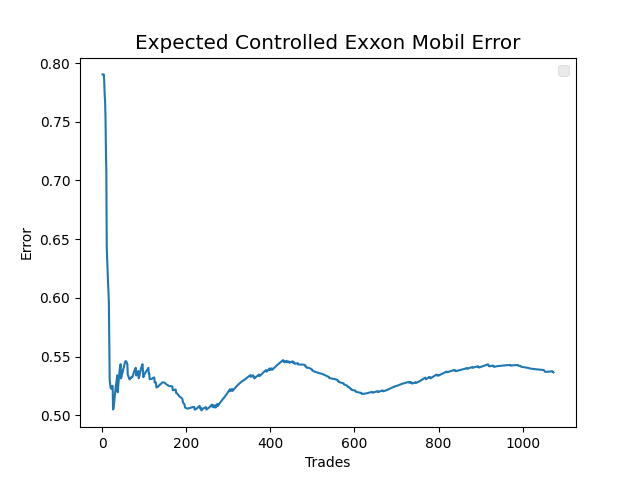}
            \caption[]%
            {{\small Exxon Mobil Controlled Error on Sept. 8, 2017}}    
            \label{fig: mean and std of net64}
        \end{subfigure}
    \end{figure*}

\begin{figure*}
         \caption[ The Expected Shortfalls for Apple, Google,  IBM, ATT, Verizon, and Exxon Mobil on Sept. 8, 2017 ]
        { \small The Expected Shortfalls for Apple, Google,  IBM, ATT, Verizon, and Exxon Mobil on Sept. 8, 2017} 
        \centering
        \begin{subfigure}[b]{0.45\textwidth}
            \centering
            \includegraphics[width=\textwidth]{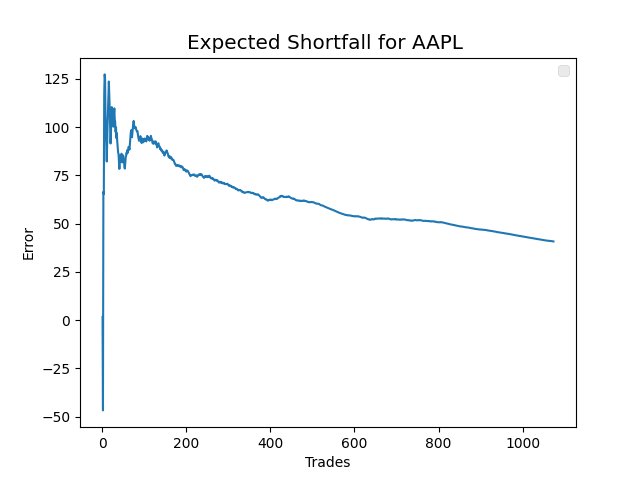}
            \caption[Network2]%
            {{\small Apple Shortfall on Sept. 8, 2017}}    
            \label{fig: mean and std of net14}
        \end{subfigure}
        \hfill
        \begin{subfigure}[b]{0.45\textwidth}  
            \centering 
            \includegraphics[width=\textwidth]{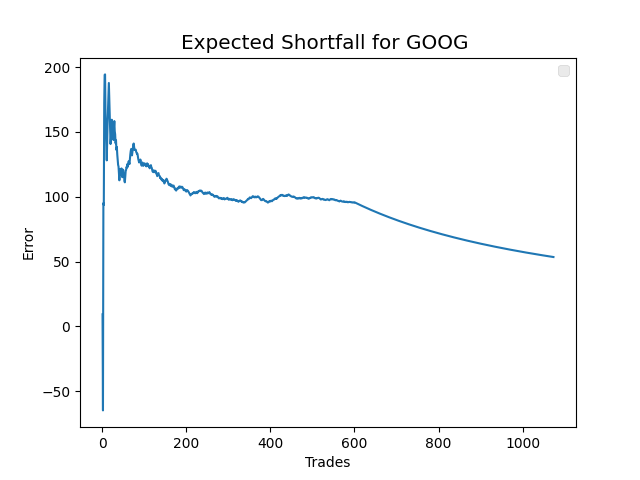}
            \caption[]%
            {{\small Google Shortfall on Sept. 8, 2017}}    
            \label{fig: mean and std of net24}
        \end{subfigure}
        \vskip\baselineskip
        \begin{subfigure}[b]{0.45\textwidth}
            \centering
            \includegraphics[width=\textwidth]{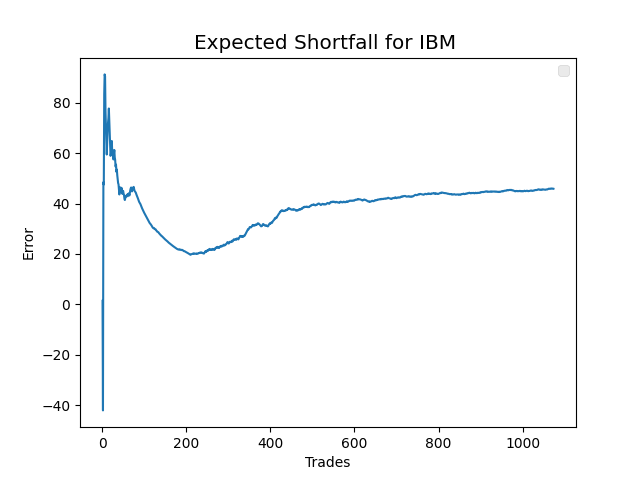}
            \caption[Network2]%
            {{\small IBM Shortfall on Sept. 8, 2017}}    
            \label{fig: mean and std of net1=34}
        \end{subfigure}
        \hfill
        \begin{subfigure}[b]{0.45\textwidth}  
            \centering 
            \includegraphics[width=\textwidth]{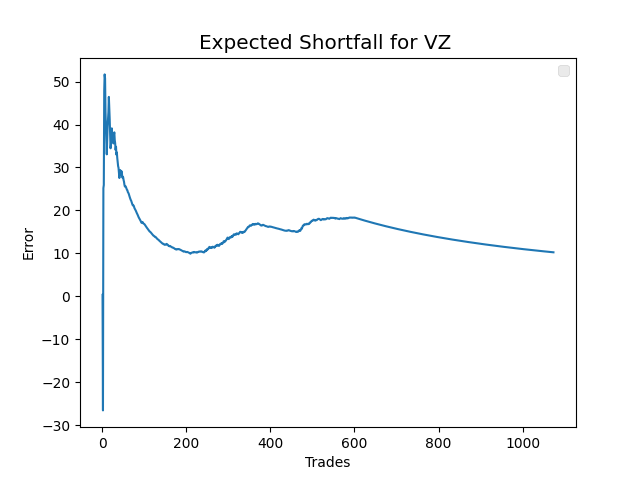}
            \caption[]%
            {{\small VZ Shortfall on Sept. 8, 2017}}    
            \label{fig: mean and std of net44}
        \end{subfigure}
        \vskip\baselineskip
        \begin{subfigure}[b]{0.45\textwidth}   
            \centering 
            \includegraphics[width=\textwidth]{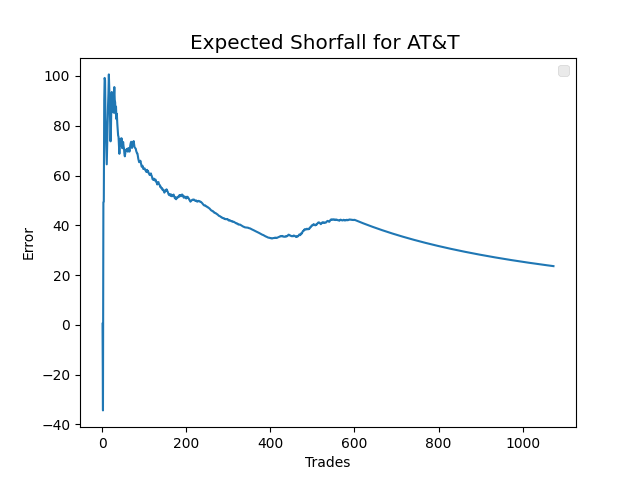}
            \caption[]%
            {{\small ATT Shortfall on Sept. 8, 2017}}    
            \label{fig: mean and std of net54}
        \end{subfigure}
        \hfill
        \begin{subfigure}[b]{0.45\textwidth}   
            \centering 
            \includegraphics[width=\textwidth]{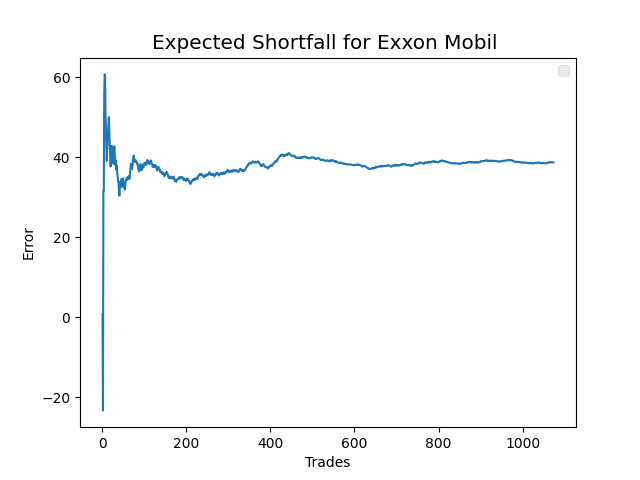}
            \caption[]%
            {{\small  Exxon Mobil Shortfall on Sept. 8, 2017}} 
            \label{fig: mean and std of net64}
        \end{subfigure}
\end{figure*}

\begin{figure*}
         \caption[The  Tracking Error for Apple, Google,  IBM, ATT, Verizon, and Exxon Mobil at Certain Action on Sept. 8, 2017 ]
        { \small The Tracking Error for Apple, Google,  IBM, ATT, Verizon, and Exxon Mobil at certain action on Sept. 8, 2017} 
        \centering
        \begin{subfigure}[b]{0.45\textwidth}
            \centering
            \includegraphics[width=\textwidth]{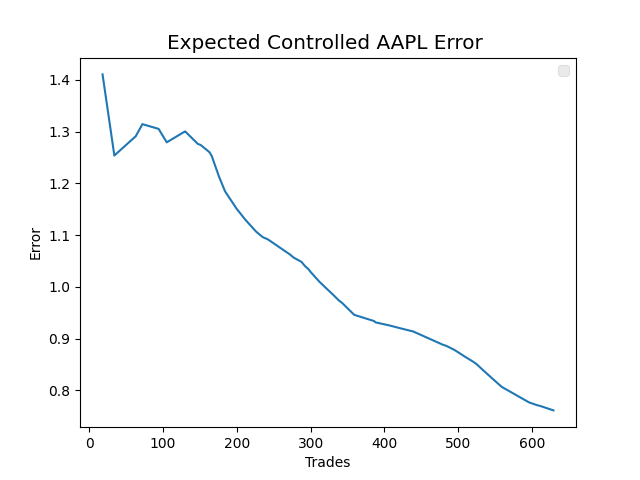}
            \caption[Network2]%
            {{\small Apple Controlled Error when Action =20 on Sept. 8, 2017}}    
            \label{fig: mean and std of net14}
        \end{subfigure}
        \hfill
        \begin{subfigure}[b]{0.45\textwidth}  
            \centering 
            \includegraphics[width=\textwidth]{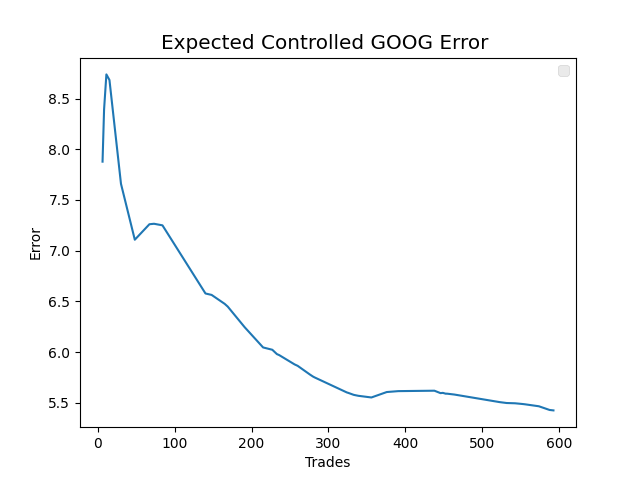}
            \caption[]%
            {{\small Google Controlled Error when Action =20 on Sept. 8, 2017}}    
            \label{fig: mean and std of net24}
        \end{subfigure}
        \vskip\baselineskip
        \begin{subfigure}[b]{0.45\textwidth}
            \centering
            \includegraphics[width=\textwidth]{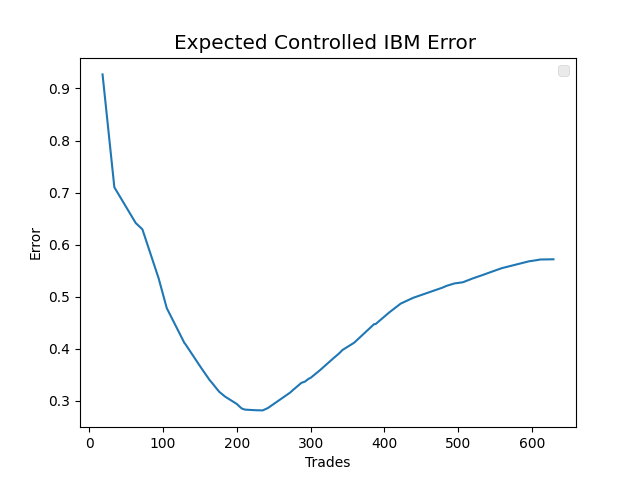}
            \caption[Network2]%
            {{\small IBM Controlled Error when Action =20 on Sept. 8, 2017}}    
            \label{fig: mean and std of net1=34}
        \end{subfigure}
        \hfill
        \begin{subfigure}[b]{0.45\textwidth}  
            \centering 
            \includegraphics[width=\textwidth]{VZShortfall.png}
            \caption[]%
            {{\small VZ Controlled Error when Action =100 on Sept. 8, 2017}}    
            \label{fig: mean and std of net44}
        \end{subfigure}
        \vskip\baselineskip
        \begin{subfigure}[b]{0.45\textwidth}   
            \centering 
            \includegraphics[width=\textwidth]{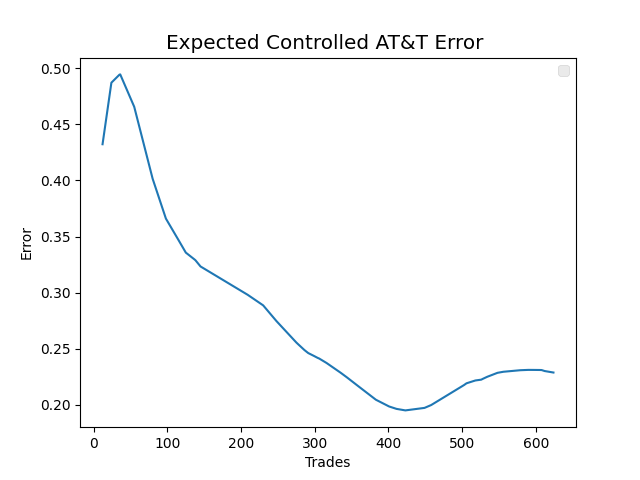}
            \caption[]%
            {{\small ATT Controlled Error when Action =100 on Sept. 8, 2017}}    
            \label{fig: mean and std of net54}
        \end{subfigure}
        \hfill
        \begin{subfigure}[b]{0.45\textwidth}   
            \centering 
            \includegraphics[width=\textwidth]{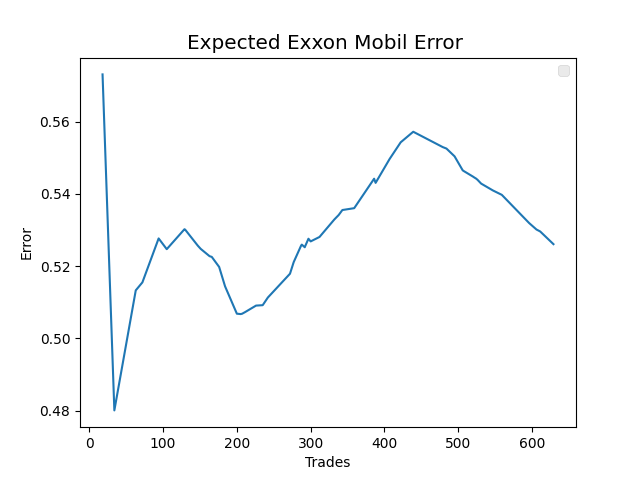}
            \caption[]%
            {{\small  Exxon Mobil Controlled Error when Action =20 on Sept. 8, 2017}} 
            \label{fig: mean and std of net64}
        \end{subfigure}
\end{figure*}

\section{Conclusion}
In conclusion, we address the challenging problem of liquidating large quantities of highly correlated stocks, a common issue for institutional investors and proprietary trading firms.  Traditional approaches to this problem are hindered by the curse of dimensionality and other stability issues, making them impractical for high-dimensional settings.  In contrast, in this paper we show how artificial intelligence, i.e., reinforcement learning, can efficiently tackle this multidimensional problem. We propose a trading algorithm that conjointly trains two neural networks to understand the intricacies and interdependencies of a basket of stocks to optimize the liquidation process and minimize the overall execution shortfall. We rigorously establish the convergence of our optimal trading strategy and demonstrate its practical application through an implementation using intra-day market data.


\newpage

\begin{thebibliography}{12}

\bibitem{AlChr1}
Almgren, R. and  Chriss, N.(1999), Value under liquidation, {\it Risk, } {\bf 12}.


\bibitem{AlChr3}
Almgren, R. and  Chriss,  N. (2000), Optimal execution with nonlinear impact functions and trading-enhanced risk, {\it Applied Mathematical Finance}, {\bf 10},  pp 1-18.

\bibitem{AlgLor1}
Almgren, R. and Lorenz, J. (2006), Bayesian adaptive trading with a daily cycle, {\it  Journal of Trading}, {\bf 1}, 4,pp.38-46.  

\bibitem{AlgLor2}
Almgren, R. and  Lorenz, J. (2007). Adaptive arrival price, {\it  Journal of Trading}, {\bf 2007}, 1,  pp. 59-66. 

\bibitem[]{AL2020}
Angoshtari, B. and Leung, T. (2020), Optimal trading of a basket of futures contracts. {\it Ann Finance }, {\bf 16}, 253–280 (2020). https://doi.org/10.1007/s10436-019-00357-w


\bibitem{Bhatnagar}
Bhatnagar, S.(2010), An Actor-critic algorithm with function approximation for discounted cost constrained Markov decision processes, {\it Systems \& Control Letters} 59, pp. 760-766.

\bibitem{BellMan}
Bellman, R. (1971) {\it Introduction to the mathematical theory of control processes},
Vol 2, Academic Press, New York and London.


\bibitem{EnFer}
 Engle, R. and  Ferstenberg, R. (2007), Execution risk: it's the same as investment risk. {\it J. Portfolio Management}, {\bf 33}, 2, pp 34-44.





\bibitem{Helmes}
 Helmes, K. (2004), Computing optimal selling rules for stocks using linear
programming, {\it Mathematics of Finance},
G. Yin and Q. Zhang (Eds),
American Mathematical Society,  pp. 87-198.



\bibitem{Marbach2001} Marbach, P., and Tsitsiklis, J. (2001), Simulation-based optimization of Markov reward processes, {\it IEEE Transaction on Automatic Control}, 46, 191-209. 

\bibitem{Pemy12} 
Pemy, M. (2012), Optimal algorithms for trading large positions, {\it Automatica}, {\bf 48}, Issue 7, pp. 1353-1358.

\bibitem{PemyZ} 
Pemy, M. (2021), Execution Shortfall Algorithms under Regime Switching, {\it Proceedings of the Conference on Control and its Applications}, Publisher SIAM, pp 48-54.

\bibitem{PemyZ1} 
Pemy, M. (2021), Optimal VWAP Strategies under Regime Switching, {\it Proceedings of the 55th  Annual Conference on Information Sciences and Systems (CISS 2021)}, Publisher IEEE.

\bibitem {Pemy}
Pemy, M. (2022), Optimal Trading Algorithms under Regime Switching, {\it The Journal of Financial Data Science} Spring 2022

\bibitem{PemyZY1}
 Pemy, M.,  Yin, G. and  Zhang, Q. (2007), Selling a large position: a stochastic control approach with state constraints, {\it Communications in Information and Systems}, {\bf 7}, No. 1, pp. 93-110.


\bibitem{PemyZY2}
 Pemy, M., Yin, G. and  Zhang, Q. (2008), Liquidation of a large block under regime switching, {\it Mathematical Finance: An international Journal of Mathematics, Statistics and Financial Economics}, {\bf 18}, Issue 4,  pp. 629-648.  



\bibitem{Sutton2000} Sutton, R. S., McAllester, D., Singh, S, and Mansour, Y. (2000), Policy gradient methods for reinforcement learning with function approximation, {\it Advances in Neural Information Processing Systems}, {\bf 12}, pp. 1057-1063.

\bibitem{Zhang}
 Zhang, Q. (2001),
Stock trading: An optimal selling rules,
{\it SIAM J. Control Optim.}, {\bf 40},  pp. 4-87.


\end{thebibliography}
\end{document}